\documentclass[10pt,journal]{IEEEtran} 
 
\usepackage[dvips]{graphicx}
\graphicspath{{../eps/}{../pdf/}{../jpeg/}}
\DeclareGraphicsExtensions{.eps,.pdf,.jpeg,.png}
\usepackage[cmex10]{amsmath} 

\usepackage{xcolor,amsthm, amsfonts, mdwmath, mdwtab, eqparbox, psfrag, setspace}
\usepackage{bbm}
\usepackage[tight,footnotesize,sf,SF]{subfigure} 
\usepackage{cite,multicol, blindtext,overpic,url,color,tikz,lipsum}
\definecolor{dg}{rgb}{0.1,0.55,0.15}

\usepackage{algpseudocode,algorithm}
\newtheorem{proposition}{Proposition}
\newtheorem{lemma}{Lemma}

\newtheorem{remark}{Remark}

\usepackage{cancel,fancyhdr,acronym}
\usepackage[T1]{fontenc}
\usepackage[normalem]{ulem}
\usepackage{soul}
\usepackage{algpseudocode}

\newcommand\bout{\bgroup\markoverwith{\textcolor{blue}{\rule[.5ex]{2pt}{1.0pt}}}\ULon}
\definecolor{dg}{rgb}{0.48, 0.25, 0.0}

\usepackage{subfigure, mathtools}
\usepackage{authblk}

\begin{document}
 
	\title{Quantum bandit with amplitude amplification exploration in an adversarial environment}

	\author{\IEEEauthorblockN{Byungjin Cho, Yu Xiao, Pan Hui \emph{Fellow, IEEE}, and Daoyi Dong \emph{Fellow, IEEE}} \vspace*{-.625432080598cm}\thanks{This work has received funding from Academy of Finland under grant
number 317432 and 318937.

B. Cho and Y. Xiao are affiliated with Information and Communications Engineering at Aalto University, Finland.
  P. Hui is with Computational Media and Arts at The Hong Kong University of Science and Technology (Guangzhou), China, Emerging Interdisciplinary Areas at The Hong Kong University of Science and Technology, Hong Kong, and Computer Science at University of Helsinki, Finland.  
 D. Dong is with Engineering and Information Technology at University of New South Wales, Australia. 
  (e-mail: \{byungjin.cho, yu.xiao\}@aalto.fi; panhui@ust.hk; d.dong@adfa.edu.au)} }

	\maketitle 
		
	\begin{abstract}
		The rapid proliferation of learning systems in an arbitrarily changing environment mandates the need to manage tensions between exploration and exploitation. This work proposes a quantum-inspired bandit learning approach for the learning-and-adapting-based offloading problem where a client observes and learns the costs of each task offloaded to the candidate resource providers, e.g., fog nodes. In this approach, a new action update strategy and novel probabilistic action selection are adopted, provoked by the amplitude amplification and collapse postulate in quantum computation theory. We devise a locally linear mapping between a quantum-mechanical phase in a quantum domain, e.g., Grover-type search algorithm, and a distilled probability-magnitude in a value-based decision-making domain, e.g., adversarial multi-armed bandit algorithm. The proposed algorithm is generalized, via the devised mapping, for better learning weight adjustments on favorable/unfavorable actions, and its effectiveness is verified via simulation.
	  
	\end{abstract}
	\begin{IEEEkeywords}
		Quantum amplitude amplification, multi-armed bandit
	\end{IEEEkeywords}%
 
 \vspace{-.345cm}

	\section{Introduction} 
	 	
	Fog computing domains, such as vehicular networks, have been rapidly proliferated \cite{Mao2017}. Enabling such emerging applications to work in a pervasive uncertain environment mandates the need for intelligent decision-making (DM) to choose a suited computing server guaranteeing the quality of service, e.g., offloaded to nodes geared with powerful computing capability. To solve the provider identification problem, sequential DM has been leveraged for its ability to learn in a trial/error fashion without explicit knowledge of the environment, while facing the exploration/exploitation (ExR/ExT) dilemma \cite{Cho2021}. The exploration strategy is known as a crucial ingredient for learning-based DM: under-ExR makes the decision stick at a sub-optimal strategy, while over-ExR may incur an ExR cost.
	 
	Various exploration strategies have been introduced to address the balancing issue, which can be categorized into three main methods of selecting an action, e.g., a service provider: i) {\color{black}An} upper-confidence bound (UCB)-type strategy, referred to as interval-estimation method \cite{Kaelbling1993}, selects an action that has the highest estimated action-value plus the UCB exploration term, making it possible to play an action that was not explored sufficiently; ii) A greedy-type strategy, referred to as the semi-uniform (SU) method \cite{Auer2002}, consists of choosing a random action with $\epsilon$-frequency or choosing the action with the highest estimated mean otherwise. For the latter, the estimation is based on
	%and otherwise choosing the action with the highest estimated mean, where the estimation is based on 
	the costs observed so far; iii) A softmax-type strategy, referred to as the probability-matching (PM) method \cite{cesa2017}, chooses actions according to a Gibbs-type probability distribution reflecting how likely the actions would be optimal, with a free parameter corresponding to inverse temperature $\beta$. With careful tuning, such a UCB-type rule is asymptotically optimal for specific cost distributions but may occur after a long period of time particularly in an adversarial environment. {Using SU and PM methods requires tuning the ExR parameter, $\epsilon$ or $\beta$, vital in a varying environment but non-trivial to set in a systematic way due to lack of generality in how to adjust the factors on favourable/unfavorable actions.}
		 
	As a promising direction to overcome the difficulties of controlling the ExR factors, adopting a quantum mechanism in the field of learning algorithms has been considered. 
	Existing works in \cite{Dong2008,Dunjko2018} show that quantum learning algorithms can achieve a better ExR/ExT trade-off compared with classical learning, and learning efficiency improvement. Such quantum enhancement arises from the use of quantum subroutines such as quantum amplitude amplification (QAA) and quantum measurement (QM). QM envisions natural ExR based on the collapse postulate of quantum mechanics, which can be used for the importance-weighted Gibbs sampling without specific exploration parameters. % settings.
	QAA, a core in Grover's algorithm \cite{Grover}, updates the probability amplitudes of actions with a certain degree of importance, performed by multiple iterations, where each can be generalized to adjust weights on favorable actions. 
 
	Existing probability amplitude updating strategies \cite{Dong2008,li2020,Fakhari2013,dong2021} suffer from arbitrary phase variation and probability amplitude jumping issues. Such uncertainty attributes may bring out severe eventuality in an arbitrarily varying environment with incomplete feedback, since the probability amplitude of a sub-optimal action could be amplified by an arbitrary degree. The concerns have not been resolved due to challenges associated with i) nonexistence of one-to-one mapping between phase and probability amplitudes and ii) nonsmoothness of arbitrary cost estimates, shed lighted in this work. To the best of our knowledge, this is the first work aiming at devising a quantum-inspired learning process in an adversarial environment with limited feedback. The features of this work can be summarized as follows. 
 
    \begin{itemize} 	
    \item This work proposes a quantum exploration-based decision-making algorithm, where a novel probabilistic action selection is adopted for enhancing an adversarial multi-armed bandit (MAB) learning strategy \cite{Auer2002}, provoked by the amplitude amplification and collapse phenomenon in quantum computation theory. 

    \item This work extends non-classical learning algorithms using a fixed phase with flexible iterations \cite{Dong2008} to their counterparts, flexible phases with an iteration, in a resembling way to existing works \cite{Fakhari2013, li2020, dong2021}. Our work differs from previous works in the ways the phases are tuned, overcoming the hardness of justifying to set a free parameter. %in their works. 

    \item This work generalizes the MAB algorithm through increasing the probability amplitude of a dominant action as well as decreasing the ones of the others. This is realized by adjusting importance weights via the devised one-to-one mapping between a quantum-mechanical phase and a learning-based decision probability, which otherwise conventionally requires an extra normalization \cite{li2020}.
 
	\item This work alleviates an undesirable situation, where a suboptimal action is amplified due to uncertainty of the empirical cost estimates in an adversarial bandit setting. This is enabled by using an implicit exploration estimate process, which renders the reduction of variance and bias simultaneously and thus achieves a better ExR/ExT balance \cite{Cho2021}. Simulation results verify its effectiveness. 
\end{itemize}	
\vspace{-.246cm}

\section{Related works}
	 
	This section presents related works in the area of quantum-enhanced exploration strategy, in terms of quantum bandit problems and amplitude amplification methods. 
		
	Quantum algorithms for bandit problems have been proposed recently \cite{Casale2020, Wang2021, Wang2021_2, Rebentrost2021}. The work in \cite{Casale2020} initiated the study of quantum algorithms for best-arm identification of MAB, the research in \cite{Wang2021} proved optimal results for best-arm identification of MAB with Bernoulli's arms, and the authors in \cite{Wang2021_2} proposed quantum algorithms to find an optimal policy for a Markov decision process with quantum speedup. These algorithms investigate potential improvements in the respective multi-armed stochastic bandit problems. The stochastic model may be unrealistic in many applications: data collected in a sequence rarely satisfy the i.i.d assumption, and it would be naive to think that corruptions never occur. The work in \cite{Rebentrost2021} studied a quantum version of the Hedging algorithm, related to the adversarial model considered pessimistic in contexts where we expect learning to be reasonably possible. However, it is limited to a bandit setting. 

    Quantum algorithms with probability amplitude updating are in general supported by two different approaches \cite{Dong2008,li2020,Fakhari2013,dong2021}. One is to make use of a fixed phase with multiple Grover iterations, which however suffers from an amplitude jumping issue \cite{Dong2008} due to discrete operations. The other is to consider a varied phase with a single iteration, which however suffers from the effects of arbitrary phase variations on the amplitudes due to nonexistence of one-to-one mapping between phase rotation and probability amplitudes \cite{li2020,Fakhari2013,dong2021}. The work in \cite{Fakhari2013} considered an empirical function mapping, e.g., setting relevant free parameters manually. However, such a manual strategy is only valid when sufficient data are available, causing unreliability. The work in \cite{li2020} considered a parametric mapping that is not reliant on empirical data. However, a substantial number of function forms remain largely unexplored, and thus such parametric strategy cannot be generalized, causing incompatibility. The work in \cite{dong2021} relaxed  the limitations of both empirical and parametric approaches. However, their approach suffers from inflexibility due to non-monotonic mapping, which fails to simultaneously amplify the dominant action and attenuate others. Additionally, none of these works considers the uncertainty of the empirical costs generated in an adversarial fashion under an information-limited environment, which could increase the probability amplitude of a sub-optimal action, leading to fatal outcomes. 
    This work addresses the aforementioned limitations by introducing a novel action updating strategy. This strategy utilizes a local one-to-one mapping between available phase rotation and relative disparity learning scores for both dominant and dominated actions. This approach allows for the simultaneous amplification and attenuation of probabilities. In addition, cumulative learning scores are used in conjunction with an implicit exploration-based biased cost estimation. This technique effectively mitigates the uncertainty associated with importance-weighted estimators in adversarial environments.
               
   \vspace{-.246cm}

 %%%%%%%%%%%%%%%%%%%%%%%%%%%%%%%%%%%%%%%%%%%%%%%%%%%
	\section{System model and learning strategy}
	This section demonstrates the system model and learning-based decision-making, applicable to offloading services. %\vspace{-0.31cm}
	\vspace{-.246cm}
	\subsection{System model}
	A service client (SC) generates tasks, while a set of service providers (SPs) $k \in \mathcal{K} = \{1,...,K\}$ execute the requested tasks with their own available resources. An SC can send a task, e.g., offloading a computational task \cite{Cho2021}, $t$ to any SP $k$ among the set. 
    Each task, $t$, is considered as a basic unit for offloading. The demand for resources from each SC may vary depending on the nature of performed applications, expressed as the multiplication of the input size $q^t$ (bits/task) and the computational complexity (cycles/bit). The service capability of an SP $k$ depends on its resource availability (cycles/sec).  The achievable up/down-link transmission rates between an SC and an SP are determined by the wireless medium characteristics. The cost for offloading a task, $D_k^t$, includes the cost for uploading the input to an SP $k$, and the execution cost at the SP, downloading the result to the SC.  
	 
	This work defines the unit service cost reflecting the service capability of each candidate SP $k$, e.g., the cost of processing one bit of input data for task $t$ on SP $k$, as $l_k^t = D_k^t/q^t$. One aim of this work is to minimize the average unit cost by optimizing the SP selection for each task in each round, $k_t$. We design a learning-based {task offloading (TO)} algorithm minimizing the expectation of the unit cost, formulated as $\mathcal{P}: \min_{k_1, k_2, \cdots, k_T} \mathbb{E}\left[\sum_{t=1}^T l_{k_t}^t\right]$, where $\mathbb{E}\left[\cdot\right]$ is the expectation, $l_{k_t}^t$ is a sequence of unit cost for the $t$-th task in the task set $\mathcal{T}$, and $T = |\mathcal{T}|$ is the number of tasks. The significance of a learning algorithm depends on the adopted benchmark policy which the algorithm is measured against. The learning regret measuring how much the SC regrets choosing its pulled action-sequence over the one with the optimal policy, is expressed as $\bar{L}_{k'}^T - \bar{L}_{k^*}^T$, where $\bar{L}_{k'}^T = \mathbb{E}\left[\sum_{t=1}^T l_{k'}^t\right]$ and $\bar{L}_{k^*}^T = \mathbb{E}\left[\sum_{t=1}^T l_{k^*}^t\right]$ correspond to the expected cumulative costs incurred by an algorithm and the optimal solution $k^* = \arg\min_k \sum_{t=1}^T l_{k}^t/T$. 
		 
	\begin{figure}[t]  		%\vspace{-15pt} 
		\centering
		\noindent\includegraphics[scale=0.385135]{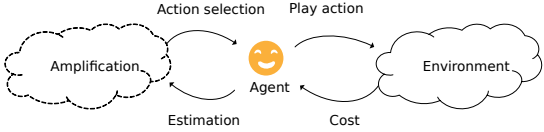}
		\vspace{-1.0em}
		\caption{System model}
		\label{system}\vspace{-1.50em}
	\end{figure}
	
	\subsection{Online learning decision-making in bandit setting} 
    Consider a framework of online learning where an SC selects one SP, $k\in\mathcal{K}$ based on an unknown cost function. There exists a trade-off between exploiting the experiential best SP for instantaneous costs and exploring the other SPs for potential benefits. The trade-off is formulated as a MAB problem specified by $\mathcal{K}$ and $l_{k}^t, t\in\mathcal{T}$.
	In an adversarial MAB, randomized policy is used such that an SC draws an arm according to a probability distribution, $k' \sim p^t = [p_{k}^t]_{k\in \mathcal{K}}$. One may employ weighted-average randomized strategy with potentials to achieve a cumulative cost as small as that of the best action \cite{Bubeck2012}. An arm $k$ is assigned with the selected probability for task $t$, $p_{k}^t$ proportional to weighted accumulated cost caused by that arm in the past, $p_{k}^t \!=\! \frac{\mathcal{W}_{k}^t}{\sum_{k} \mathcal{W}_{k}^t}$ where $\mathcal{W}_{k}^t$ is a weight of each arm $k$. A score-based learning process is considered as follows: service capability of an SP can be represented by a score, cumulative per-bit cost up to $t-1$, $\mathcal{\hat{L}}_{k}^{t-1} \!\!=\!\! \sum_{t'=1}^{t-1} \eta_{t'} \hat{l}_{k}^{t'}$, where $\hat{l}_{k}^{t'}$ is the cost estimate from the arm $k$ for task $t$ and $\eta_{t'}\!\in\! (0,1]$ is the learning rate. Considering exponential potential with the score, $\mathcal{W}_{k}^t \!=\! e^{-\mathcal{\hat{L}}_k^{t-1}}$, the importance-weighted mechanism assigns exponentially higher probability to strategy with lower cumulative scores up to $t\!-\!1$ due to $\frac{\partial p^t}{\partial L_k} \!<\!0$ where $L_k \!=\! \mathcal{\hat{L}}_k^{t-1}$. The scores reinforce the success of each strategy measured by the estimated TO cost, so an SC would rely on the strategy with the lowest one. 
	
	\vspace{-.0506cm}
	
	%%%%%%%%%%%%%%%%%%%%%%%%%%%%%%%%%%%%%%%%%%%%%%%%%%%
	\section{Quantum amplification exploration strategy}
	We develop a quantum learning-based TO algorithm, enabling an SC to learn the TO costs of candidate SPs and to choose an SP in aid of quantum subroutines.%, QM and QAA. 	
		\vspace{-.26cm}
 
	\subsection{Learning system with quantum concepts}
	An action in a learning system is represented with a quantum state, inspired by the advantages of quantum computation. Prior to the action selection carried out by observing the state according to collapse postulate of QM, the state specified by probability amplitude is updated by a QAA process.

	\subsubsection{Quantum basics} 
	The fundamental information unit in quantum computation is the quantum bit (qubit). A qubit denoted as $|0\rangle$ and $|1\rangle$ corresponds to the states $0$ and $1$ for a classical bit. Also, a qubit can lie in both $|0\rangle$ and $|1\rangle$ at the same time, a linear combination of $|0\rangle$ and $|1\rangle$, expressed as $|\Psi\rangle = g_0|0\rangle + g_1|1\rangle$ where $g_0$ and $g_1$ are complex coefficients. This quantum phenomenon is called state superposition principle. 
	 When we measure a qubit in superposition $|\Psi\rangle$, the qubit system would collapse into one of its basic states $|0\rangle$ with probability $|g_0|^2$ or $|1\rangle$ with probability $|g_1|^2$. Thus, $g_0$ and $g_1$ are in general called probability amplitudes whose magnitude and argument represent amplitude and phase, respectively, satisfying $|g_0|^2 + |g_1|^2 = 1$.		
	According to quantum computation theory, a fundamental
	operation in the quantum computing process is a unitary transformation $U$ on the qubits. If one applies a transformation $U$ to a superposition state, the transformation will act on all basis vectors of this state and the output will be a new superposition state obtained by superposing the results of all basis vectors. The transformation can simultaneously evaluate the different values of a function for a certain input and it is called quantum parallelism.   
 
 \subsubsection{Collapsing action selection} 
	 
	A quantum state $|\Psi\rangle$ can describe the state of a quantum system. %, which is a unit vector, i.e., $\langle\Psi|\Psi\rangle = 1$ in Hilbert space. 
	The work in \cite{Dong2008} proposed a formal representation for the quantum system with multiple actions. Let $K$ be the number of actions, %then choose number of $n$ characterized by e.g., 
	$K= 2^n$ 
	%the following inequality, $K\leq 2^n \leq 2K$ 
	where $n$ qubits are used to represent eigenactions\footnote{The actions in the classical system are denoted as the corresponding orthogonal bases and are called the eigenactions in a quantum system.}. For an $n$-qubit system, its quantum state can be represented with tensor product of $n$ independent qubits $|\Psi\rangle = |\Psi_1\rangle\otimes |\Psi_2\rangle\otimes\cdots \otimes |\Psi_n\rangle$ where $\otimes$ means tensor product and $|\Psi_v\rangle$ represents the $v$-th ($v\!\in\![1, n]$) qubit in the superposition state of $|0\rangle$ and $|1\rangle$. 
	According to \cite[Prop.1]{Dong2008}, %an arbitrary state in quantum bandit learning can be expanded in terms of an orthogonal set of eigenactions. 
		{for an $n$-qubit learning system, its quantum state at $t$ can be expressed as $|\Psi^t\rangle = \sum_{a\in \mathcal{A}^t} {g_a^t}|a\rangle$ where $\mathcal{A}^t$ is the set of $2^n$ eigenactions, each of which with $n$ length of a binary string, and ${g_a^t}$ is the complex coefficient, the probability amplitude}\footnote{Amplitudes correspond quantum probabilities representing the chance that a quantum state will be collapsed to when being observed.} of eigenaction $|a\rangle$ subject to $\sum_{a\in \mathcal{A}^t} |{g_a^t}|^2 = 1$. The index $t$ is omitted below for ease of description. The quantum representation establishes a bridge between the eigenactions $\mathcal{A}$ and the arms $\mathcal{K}$, shown by 
		%	 	\begin{IEEEeqnarray}{llllll} %\label{grover_update}
			$|\Psi\rangle \!\!= \sum_{a\in \mathcal{A}}  {g_a}|a\rangle \!\rightarrow\! \sum_{k\in \mathcal{K}} {g_k}|k\rangle$.
			%	 	\end{IEEEeqnarray}
		The actions can be represented by $\log_2\!K$ qubits, denoted by $|1\rangle,\! \cdots\!, \!|{K}\rangle$. An SP selected by an SC before any QM is {implemented on} a superposition state $|\Psi\rangle$ which would collapse to one of its eigenactions with probability {$p_k = |{g_k}|^2$}, $|\Psi\rangle \!\rightarrow \!|k\rangle$ when an agent measures the quantum state according to the collapse postulate of quantum mechanics \cite{Dong2008}. Such quantum collapse phenomenon can be considered as creating information on action selection strategy, e.g., $k'\sim p$ where $p = [p_1,\cdots, p_{{K}}]$.  
	  
		\subsubsection{Amplifying probability amplitude} 
		Before the collapse, the probability amplitudes of eigenactions can be reshaped via a QAA subroutine, e.g., Grover iterations, each of which gradually modifies the collapsing probabilities. The evolution of a system is described by a unitary transformation performed on the superposition states of its possible eigenactions to amend the probability amplitudes updated after $n$-Grover iterations on $|\Psi_{0}\rangle$, a state before amplification, viewed as
		\begin{IEEEeqnarray}{llllll} \label{grover_update}
			|\Psi^{}\rangle = G^n \cdot |\Psi_{0}\rangle,
		\end{IEEEeqnarray}
		%	where $|\Psi_0\rangle = \sum_{k\in \mathcal{K}} \sqrt{\mathcal{W}_k/\sum_k \mathcal{W}_k}|k\rangle$ 
		where {$|\Psi_0\rangle = \sum_{k\in \mathcal{K}} g_k|k\rangle$} and $G$ is a Grover iteration which has two substeps, an oracle query and a diffusion operation, built in a form of the unitary as follows
		\begin{IEEEeqnarray}{llllll} \label{grover_update2}
			G = -U_{(\phi_2,\Psi_0)} \cdot U_{(\phi_1,m)}, 
		\end{IEEEeqnarray}
         where $U_{(\phi_1,m)}$ is an operation based on an oracle query, shifting the phase of the target action\footnote{Classically, $m = \arg\max_k {p}_k$, while non-classically done by \cite{DHH}.} $|m\rangle$ with $\phi_1$, and $U_{(\phi_2,\Psi_0)}$ is a diffusion operation, rearranging the phases of all actions with $\phi_2$. The two unitary operators, employed for the targeted action $|m\rangle$ before amplification, {$|\Psi_0\rangle = {g_m}|m\rangle + g_{\breve{m}}|\breve{m}\rangle$} where $|\breve{m}\rangle$ is the vector orthogonal to $|{m}\rangle$, are expressed as $U_{(\phi_1,m)} = I - (1-e^{j\phi_1})|m\rangle \langle m|$ and $U_{(\phi_2,\Psi_0)} = I - (1-e^{j\phi_2})|\Psi_0\rangle \langle \Psi_0|$ where $I$ is the identity matrix, $\langle m|$ and $\langle \Psi_0|$ are Hermitian transposes of $|m\rangle$ and $|\Psi_0\rangle$. While two operators have no effect on $\breve{m}$ except normalization, they amend the target action's amplitude. %\vspace{-.2cm}

	\subsection{Quantum amplitude amplification based exploration}
	
	%\subsection{Quantum-enhanced exploration strategy}
	The effect of the Grover iterations on $ |\Psi_0\rangle$, due to its probability updating nature, can be used as a quantum learning strategy. A natural question is how to amplify/attenuate the amplitudes appropriately, yielding a better exploration strategy.

	\subsubsection{Controlling probability amplitude}
	Note that the parameters, $\phi_1$, $\phi_2$, and $n$ in (\ref{grover_update}) and (\ref{grover_update2}) determine how the probability amplitudes are updated. The transformation can be executed with proper values of the parameters corresponding to importance weights for the eigenactions. Different amplitude updating approaches have been considered in \cite{Dong2008, Fakhari2013, li2020, dong2021}. Generally, one is to fix $n \!=\! 1$ with varied values of $\phi_1$ and $\phi_2$ as learning-related factors, and another is to use a feasible value of $n$ with fixed values of $\phi_1$ and $\phi_2$. Since the latter suffers from intermittent update in the amplitudes, the former is adopted in this work, i.e., $n \!=\! 1$ with varied $\phi_1$ and $\phi_2$.  
	\begin{lemma} \label{lemma1}(Impact of G)
		The updated coefficients in amplification/attenuation, defined as the ratio between the amplitudes of targeted/untargeted actions, after being acted by an operator $G$ and before that, can be expressed as $\varrho$ and $\varsigma$ where
		\begin{IEEEeqnarray}{llllll}
			\varrho &=  |(1-e^{j \phi_1}-e^{j \phi_2}) - (1-e^{j \phi_1})(1-e^{j \phi_2}) p_m|^2 ~\mbox{and}~\nonumber\\
			\varsigma &= |-e^{j \phi_2} - (1-e^{j \phi_1})(1-e^{j \phi_2})p_m|^2.\nonumber
		\end{IEEEeqnarray}
 
		\begin{proof}
			After applying one operator $G$	on $|\Psi_0\rangle$, the amplitude vector in the next iteration becomes $|\Psi\rangle = G |\Psi_0\rangle$ shown as $G|\Psi_0\rangle = (P-e^{j\phi_1}){g_m}|m \rangle + (P-1) g_{\breve{m}}|\breve{m}\rangle$, where $P = (1-e^{j\phi_2}) [1-(1-e^{j\phi_1}) p_m]$. The updated probabilities of the selected and unselected actions, $|m\rangle$ and $|\breve{m}\rangle$, can be obtained by $\varrho \cdot p_m$ and $\varsigma \cdot p_{\breve{m}}$ where the ratios of the amplitudes between after and before $G$ are $\sqrt{\varrho} = 1-e^{j \phi_1}-e^{j \phi_2} - (1-e^{j \phi_1})(1-e^{j \phi_2})p_m$ and $\sqrt{\varsigma} = -e^{j \phi_2} - (1-e^{j \phi_1})(1-e^{j \phi_2})p_m$ \cite{dong2021}.
		\end{proof}
	\end{lemma}

	\subsubsection{Mapping phase/probability amplitudes}
	Note that the overall effect of $G$ on $ |\Psi_0\rangle$ is a two-substep phase rotation amplitude enabling to update probability amplitude, i.e., by selecting feasible $\phi_1$ and $\phi_2$, it is possible to manipulate the values of $\varrho$ and $\varsigma$. 
	While existing works in \cite{Fakhari2013, li2020, dong2021} focused on updating the probability amplitude of a target action only, e.g., amplifying/attenuating the amplitude for a good/bad action, they {have limited capability} of generalizability and complexity: requiring i) a free parameter selection indicating an amplified/attenuated degree but {varying} for different situations and ii) a re-normalization updating probability amplitudes of untarget actions, both of which are due to lack of one-to-one mapping between quantum probability and phase rotation amplitudes. This work proposes a pipeline to support the mapping operation by designing a local monotone function.

	\begin{lemma} \label{lemma_phi_opposite}
		(Impact of $\phi$) Setting $\phi =  \phi_1 =  \phi_2$ allows for updating the values of $\varrho$ and $\varsigma$ simultaneously but oppositely.	 
		\begin{proof}
			Note that two functions, $(1-\varrho)$ and $(1-\varsigma)$ have opposite signs due to the facts that i) $0<p_m < 1$, ii) $1-\varrho = (p_m-1)\kappa$ and iii) $1-\varsigma  = p_m \kappa$ where $\kappa  = 4(2 p_m - 1) \sin^2(\phi_1/2) (\cos \phi_2 - 1) + 2\sin \phi_1 \sin \phi_2$. It is straightforward to conclude that $\varrho$ and $\varsigma$ are designed to be larger or smaller than 1, respectively but conversely, irrespective of $\phi_1$ and $\phi_2$. Based on the phase matching condition \cite{Long2002}, $\phi =  \phi_1 =  \phi_2$, their second derivatives w.r.t $\phi$ also have signs opposite each other due to $\frac{\partial^2 (1-\varrho)}{\partial \phi^2} = (p_m-1) \kappa'$ and $\frac{\partial^2 (1-\varsigma)}{\partial \phi^2} = p_m \kappa'$ where $\kappa'  = (4-8 p_m)\cos\phi + 8 p_m \cos2\phi$. Such a converse relation between $\varrho$ and $\varsigma$ allows focusing on updating one of them.%, i.e., setting $\phi$ given $|h_i|^2$ results in simultaneous updates on $\varrho$ and $\varsigma$, but amplified $\varrho$ and attenuated $\varsigma$, or reverse.	
		\end{proof}
	\end{lemma} 
	\setlength{\textfloatsep}{0.1cm}
\setlength{\floatsep}{0.1cm}
	An action is rewarded/punished with higher/lower unit effort. To determine an updating degree, e.g., establishing how much it would be amplified/attenuated, the differences in learning scores between the optimal arm and sub-optimal ones can be considered, $\mathcal{D} = \mathcal{W}/\|\mathcal{W}\|_{\infty} = [e^{-(L_k-\min(L))}]_{k\in\mathcal{K}} $ where $L = [L_1, \cdots, L_K]$ and $L_k = \mathcal{\hat{L}}_k^{t-1}$,
	representing the relative disparity between targeted and untarget actions. Due to the fact that the values are lower than or equal to 1 for all actions, we map the average obtained relative disparity $\bar{\mathcal{D}}$ to the ratio $\varsigma$ via an appropriate adjustment of $\phi$. To diminish the probabilities of untarget actions proportional to $\bar{\mathcal{D}}$, one may find a range where probability amplitudes vary monotonically.

	\begin{figure}[t]
		\centering  	%	\vspace{-15pt} 
		% {\includegraphics[width=0.35\textwidth]{./code_matlab/grover-master/grover25_amplitudeRatio7_Proflies_updated.eps}} \vspace{-.851em} 
  {\includegraphics[width=0.343\textwidth]{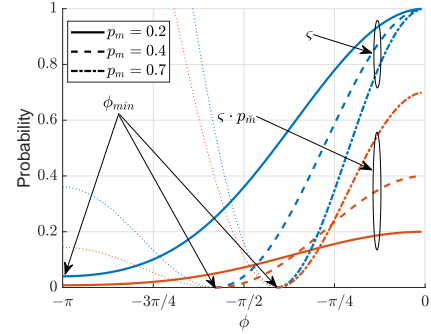}} \vspace{-.851em} 
		\caption{Profiles of $\phi$ and $\varsigma$ w.r.t $p_m$. One example of $\varsigma \cdot p_{\breve{m}}$ where $p_{\breve{m}} = p_m$.}
		\label{numerical_fig_ratio1}
		\end{figure}
	
	Next, we show how to establish $\phi$ for the amplitude amplification, by identifying local monotonic function of $\varsigma$ on $\phi$ and specifying a one-to-one mapping between $\bar{\mathcal{D}}$ to $\varsigma$.
	%$\varrho$ and $\varsigma$ can be properly updated by 
	\begin{proposition} 	\label{prop_phi}
		(Finding of $\phi$) The ratios $\varrho$ and $\varsigma$ can be controlled via a phase $\phi = -\arccos\left(W_{{(1-\varsigma_{min})\bar{\mathcal{D}}  + \varsigma_{min}}}\right)$ 
	where $W_x = 1-\left(\frac{1-\sqrt{x}}{2p_m}\right)$ and $\varsigma_{min}  = \max[(1-4p_m)^2,0]$. \vspace{-.3cm}
		\begin{proof}
			%(Local monotonicity of $\varsigma$ on $\phi$)
			Note that a ratio of $\varsigma$ is monotonically increasing within a specified range. The ratio $\varsigma$ has local maximum/minimum points at $\phi =  \left\lbrace 0, \pi, \arccos \left(1-\frac{1}{2p_m} \right) \right\rbrace$, each of which satisfying $\frac{\partial \varsigma}{\partial \phi} = 0$. And it increases in $\phi$, $\frac{\partial \varsigma}{\partial \phi} > 0$, when case i) $\sin \phi <0$ and $\cos \phi > 1 - \frac{1}{2 p_m}$, or case ii) $\sin \phi >0$ and $\cos \phi < 1 - \frac{1}{2 p_m}$ is satisfied, fulfilled with $-\operatorname{Re}\left[\arccos (W_0) \right]< \phi < -\operatorname{Re}\left[\arccos (W_1) \right]$ for case i), or $\operatorname{Re}\left[\arccos \left(W_0\right)\right]< \phi < \operatorname{Re}\left[ \arccos\left(W_{-1}\right) \right]$ for case ii), respectively, where $W_x = 1-\left(\frac{1-\sqrt{x}}{2p_m}\right)$. While for case i) a phase value of $\phi$ may have different maximum values of $\varsigma$ for different $p_m$ in its increasing range, for case ii), a ratio value of $\varsigma$ monotonically increases in $\phi$ ranged from 
			\begin{IEEEeqnarray}{llllll} \label{eq_range}
				\phi_{min} \leq \phi <  0{\color{blue},} 
			\end{IEEEeqnarray}	
			where $\phi_{min} = -\min\left[\arccos \left(W_0\right), \pi\right]$, for $\varsigma_{min}  \leq \varsigma < 1$ with $\varsigma_{min}  = \max[(1-4p_m)^2,0]$, and reaches the maximum equal to 1 only at $\phi = 0$ irrespective of $p_m$, which allows us to focus on case i), see Fig.\ref{numerical_fig_ratio1}. Note that a ratio value of $\varsigma = 1-p_m \kappa$ in Lemma~\ref{lemma_phi_opposite} increases w.r.t a phase value of $\phi = -\arccos \left(W_{\varsigma}\right)$ satisfying Eq. (\ref{eq_range}). The feasible $\phi$ is set to be proportional to the average obtained relative disparity $\bar{\mathcal{D}}$ which could be one-to-one mapped to the range of $\varsigma$ given $p_m$. Thus, the ratios, $\varrho$ and $\varsigma$, can be controlled via $\phi \!=\! -\arccos \left(W_{{(1-\varsigma_{min})\bar{\mathcal{D}}  + \varsigma_{min}}} \right)$.%, simulataneously. 
		\end{proof}
	\end{proposition}

	\begin{remark}% (Attenuation sensitivity) 
		(Profiles of $\phi$ and $\varsigma$) Note that $\varsigma$ decreases in $p_m$ due to $\frac{\partial \varsigma}{\partial p_m}<0$ in Prop.~\ref{prop_phi}, and thus attenuated probabilities are achieved, see Fig.~\ref{numerical_fig_ratio1}. 
		For a high $p_{m}$, the impact of $\phi$ on $\varsigma$ becomes large, and thus $\phi$ can be tuned within a small variation range for the updating. Contrarily, for a relatively small $p_{m}$, a much larger degree of freedom on $\phi$ adjustment is configured, a natural way to avoid local maxima with a relatively small $p_{m}$. Setting $\phi$ tunes $\varrho$ and $\varsigma$, simultaneously. %\vspace{-.42cm}
	\end{remark}

	\subsubsection{Processing implicit cost estimation}
	An SC selects an arm for a task and receives the cost from the selected arm, not from the others.
	The cost from an arm  $k \neq k'$ could not be observed due to incomplete feedback in the bandit problem. One may use an unbiased estimate, $\hat{l}_{k}^t = \frac{{l}_{k}^t\cdot \mathbbm{1}_{k = k'}}{p_{k}^t}$, but {it could cause} large fluctuation in the cost due to inverse-proportion to $p_{k}^t$. Instead, this work considers Exp3 algorithm endowed with implicit exploration (IX)-style cost estimates \cite{Neu2015}, which controls the variance at the price of extra bias. After each action, the cost estimate is calculated as $\hat{l}_{k}^t  =  \frac{{l}_{k}^t\cdot \mathbbm{1}_{k = k'}^t}{p_{k}^t + \gamma_t}$, a biased estimator due to $\mathbb{E}[\hat{l}_{k}^t] = \sum_{k } p_{k}^t \hat{l}_{k}^t  \leq l_{k}^t$, where $\gamma_t \in (0,1]$ is the implicit learning rate. While actions with large costs are set to be negligible probabilities by the classical recipe \cite{Bubeck2012}, such an implicit price allows them to have low but non-negligible ones and to be chosen occasionally. Thus, the estimator could guarantee performance with high probability.

	\begin{algorithm}[t]%\small 
		\caption{Quantum amplification exploration strategy}  \label{algorithm1}
		%\begin{minipage}{\dimexpr\linewidth-\SpaceReservedForComments\relax}
		\begin{algorithmic}[1]  
			\State Input: $\eta_t>0$, $\gamma_t>0$, $\mathcal{K}=\emptyset$, $\mathcal{W} \leftarrow \vec{1} \in \mathcal{R}^{K}$ 
%			\State Set $\mathcal{W} \leftarrow \vec{1} \in \mathcal{R}^{K}$
			\For{$t\in \mathcal{T}$}   
			\State Set $p \leftarrow \mathcal{W} /\|\mathcal{W}\|_1$  %{\color{red} new arm}
			\State {Set $|\Psi_0\rangle \leftarrow$ preparing $\sum_k g_k| k\rangle$ where $|g_k|^2 = {p_k}$}
			\State Set $|\Psi\rangle ~\leftarrow$ updating $(\varrho,\varsigma)$ with $\phi$ set by Prop. \ref{prop_phi}		
			\State Set $k'~~~ \leftarrow$ measuring $|\Psi\rangle$ and play the strategy ${k'}$% \Comment{Classic}
			\State Get ${l}_{k'}$  and update $\mathcal{W}$ with $\eta_t, \gamma_t$ {by Prop. \ref{prop_regret}}	
%			\State Get the feedback and suffer the cost ${l}_{k'}$  
%			\State Update the weights $\mathcal{W}$ with $\eta_t, \gamma_t$ {by Prop. \ref{prop_regret}}	
			\EndFor 
			\State Output: sequences $\sum_{t\in\mathcal{T}} l_{k'}^t>0$
		\end{algorithmic}	
	\end{algorithm}  
	 
	\subsection{Proposed algorithm} 
	The workflow of the proposed algorithm (Algorithm \ref{algorithm1}) can be divided into three parts: i) interaction, ii) estimation and iii) selection. While the {first} part is about a typical interaction as an external learning process, the last two parts correspond to a classical and quantum-inspired operation as an internal learning process.
	An iterative method is used to link the conventional outer and inner processes such that the classical information is conveyed from a step $t$ to the next $t+1$ via interaction between an agent and the adversary, including: strategy playing, feedback getting, and cost suffering. The internal learning process is characterized by the score updating rule, and the local selection rule defined by what action is output given the score (selection). The algorithm is designed in a modular way so that its quantum-inspired part can be treated as a separate building block where the quantum enhancement is exhibited, whose source lies in the use of quantum subroutines to perform each internal selection process.  
	The probability distributions $p^t\in\mathcal{R}^K$ are passed to the quantum subroutines where, instead of sampling one action in a classical manner, in a quantum setting, one sample can be obtained by preparing the state {$|\Psi_0\rangle = \sum_{k\in\mathcal{K}} {g_k^t} |k\rangle$ where $|g_k^t| = \sqrt{p_k^t}$, %$|\Psi_0\rangle = \sum_{k\in\mathcal{K}^t} \sqrt{p_k^t} |k\rangle$
    updating it with the proposed amplification, see Prop. 1, and measuring the updated $|\Psi\rangle$, e.g., collapsing action selection.}
	
	\begin{proposition} 	\label{prop_regret}
		%(Lower regret)
		The quantum strategy with $\phi \!\neq\! 0$ {can} achieve better regret than the one with $\phi\!\!=\! 0$, when $\eta_t\!>\frac{1}{t}\!$ and $\gamma_t\!>\!\frac{1}{2t}$.
		\begin{proof}
			Assume that a dominant arm's index is $m$, $L_{m} \leq L_{k}, \forall k\in\mathcal{K}$, one non-dominant arm selection $k \in \mathcal{K}\backslash m$
			for $t$ yields ${p_{k}^t} > p_{k|{\phi \neq 0}}^t$, while a dominiant one yields ${p_{m}^t} < p_{m|{\phi \neq 0}}^t$. Further proof is omitted, being analogous to the proof of \cite[Props. 2 and 6]{Cho2021}. 
		\end{proof}
	\end{proposition}
	
	\begin{remark}
		Note that the collapse of a quantum state is not real selection, but just a fundamental phenomenon when the state is measured, resulting in i) a good ExR/ExT balance and ii) a natural action selection without setting parameters unlike conventional approaches. The agent can explore its strategies in superposition in a way that guarantees a provable regret improvement in its learning time over its classical analogue.
	\end{remark}
	
 \vspace{-.3cm}

	\section{Performance evaluation} 
	This section conducts numerical studies to assess the learning performance of the proposed algorithm.
	
	Consider an SC, requesting the computational resource from candidate SPs. The distance between the SC and each SP is to follow a uniform distribution, $d\sim\mathcal{U}[0,d_r]$ where $d_{r}$ is the communication range equal to 400 m. The transmission power of the SC is $24$ dBm, the channel bandwidth is $10$ MHz, and noise power is $-174$ dBm/Hz, and
	large/small-scale fading gains follow $128.1 + 37.6\log_{10}(d)$ and Rayleigh distributed with unit variance, respectively. The interference effects on the co/adjacent channel are assumed to be ignored \cite{Cho2021}. Consider 5 SPs with maximum CPU frequency, $F_k \in \{6, 6, 5, 4, 3.5\}$ GHz for $T=3e3$. For an SP, the allocated CPU frequency to the SC is a fraction of the maximum distributed from 20\% to 50\%, but arbitrarily constrained \cite{Cho2021}. The computational complexity and task size are set to $1e3$ cycles/bit and $1e6$ bits/task. 
	 
	\begin{figure*}[t]
	\centering  	%	\vspace{-15pt} 
	\begin{tabular}{c c c} 
		%	\hspace{-15pt} 
		\hspace{-12pt} 		
		\subfigure[\label{numerical2:regret1}]{\includegraphics[width=0.332\textwidth]{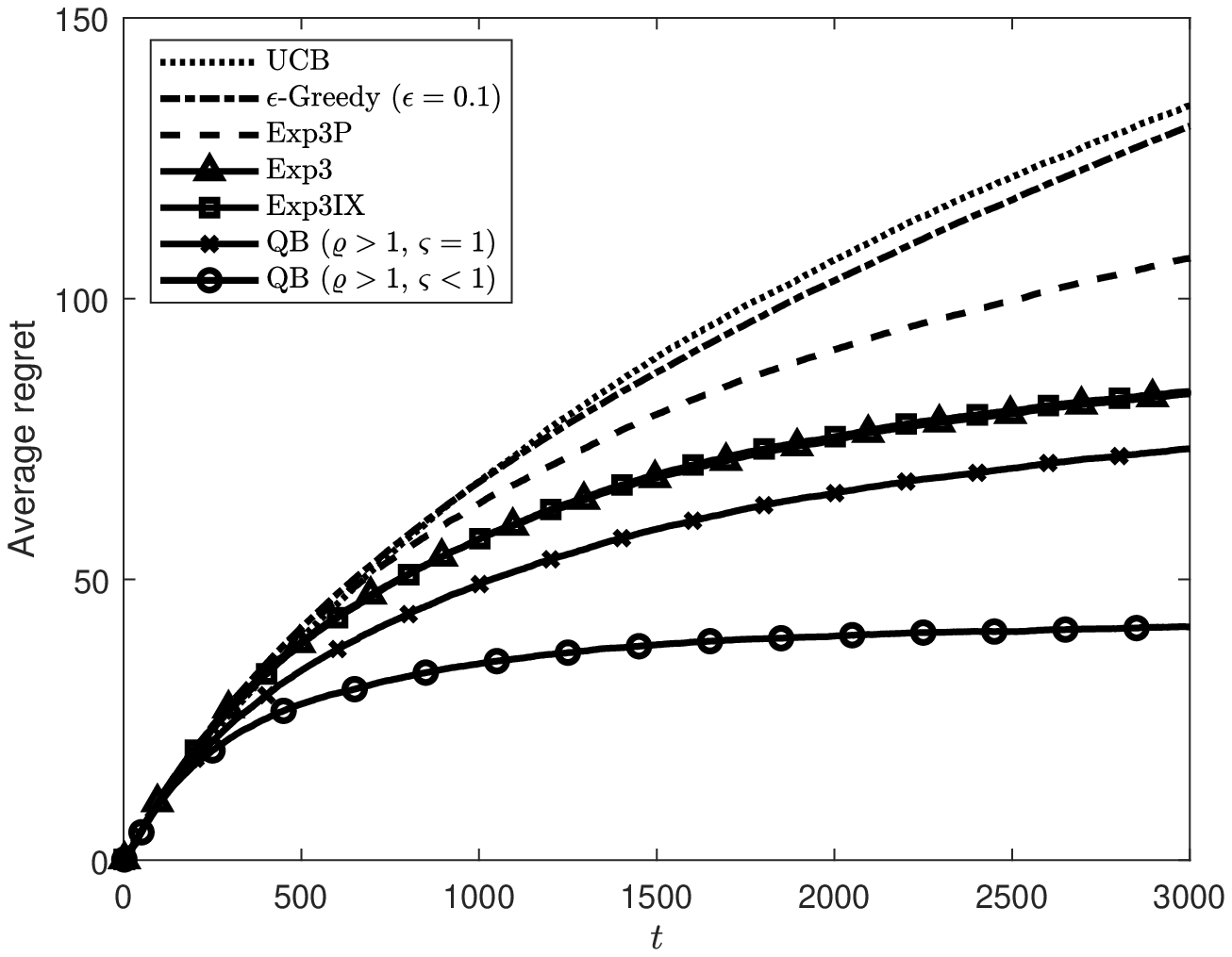}} 
		&
		\hspace{-20pt} 		
		\subfigure[\label{numerical2:regret2}]{\includegraphics[width=0.332\textwidth]{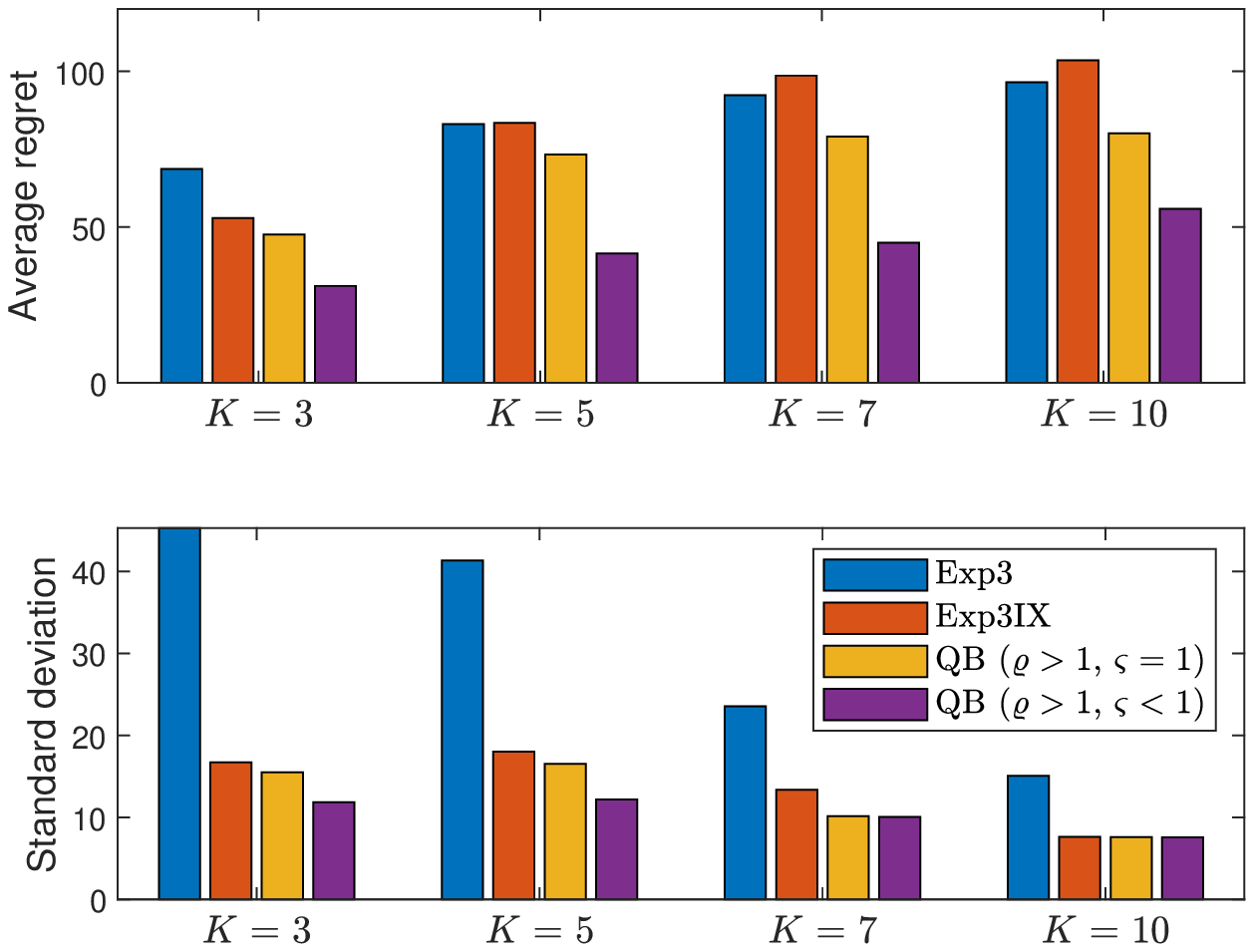}} 
		&
		\hspace{-20pt} 	
		\subfigure[\label{numerical2:fig_ratio2}]{\includegraphics[width=0.332\textwidth]{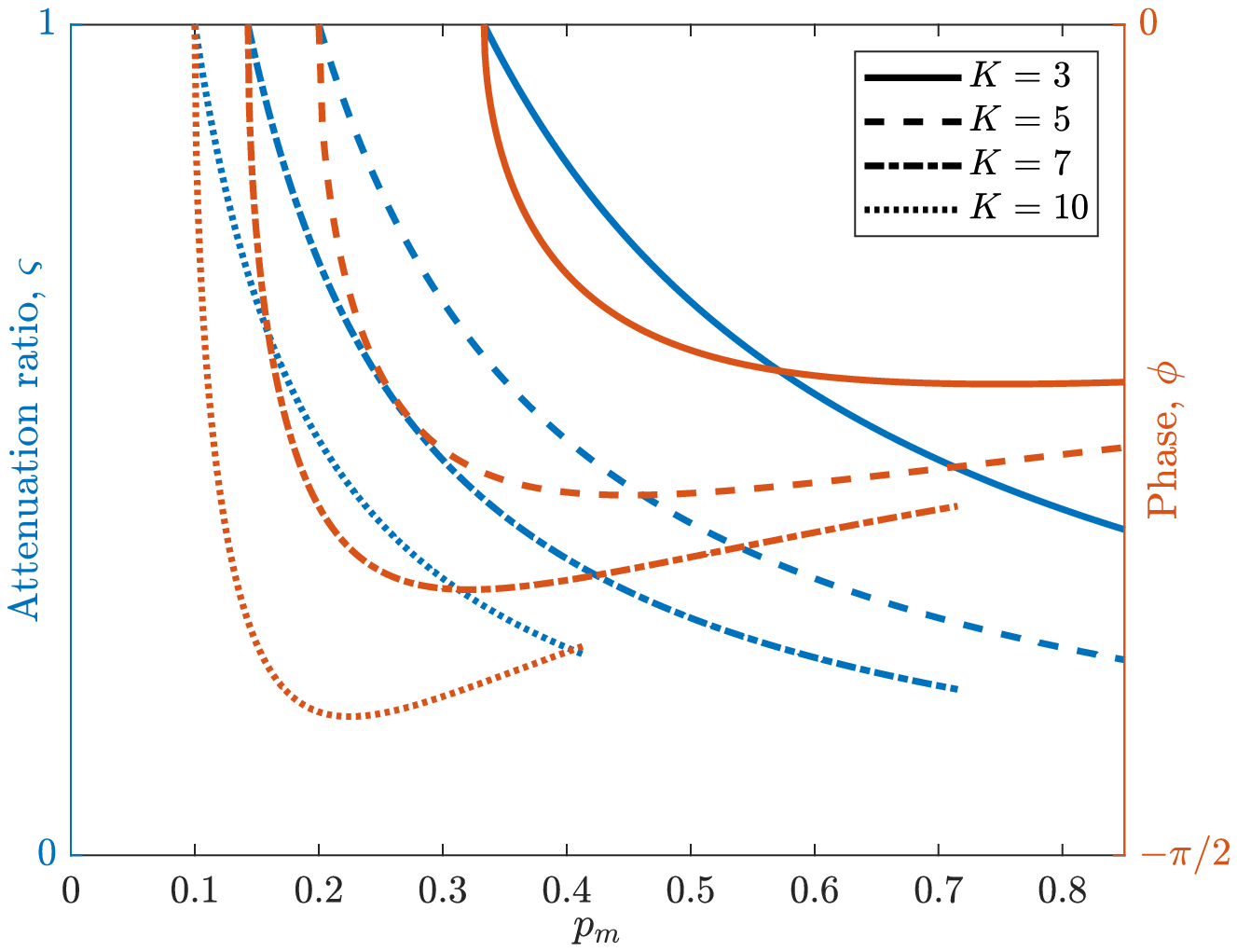}}			
	\end{tabular} \vspace{-.91em}
	\caption{(a) Regret w.r.t $t$ for $K = 5$, {(b) Regret at $T$ with $F_k \!=\! F_{mod(k,5)},\forall k\in\!\mathcal{K} = \{1,\!\cdots,\!K\}$ for different $K$, and (c) their selected $\varsigma$ and $\phi$ w.r.t $p_m$.}}
	\vspace{-.951em}
	\label{numerical2}
	%	\caption{Impact of suitable selection policy on the per-task performance (in single epoch)}\vspace{-.13em}
\end{figure*}
 
	The proposed quantum algorithm is compared to the conventional counterparts in terms of the learning regret. Those counterparts include choosing arms based on i) upper confidence bound such as UCB~\cite{Auer2002-2}, ii) current knowledge with a probability $1-\epsilon$ such as $\epsilon$-Greedy~\cite{Bubeck2012} when $\epsilon = 0.1$, and iii) probability matching such as Exp3~\cite[Sec 3.1]{Bubeck2012} guaranteeing an expected regret bound, Exp3P~\cite[Sec 3.2]{Bubeck2012} and Exp3IX~\cite{Neu2015} guaranteeing a high probability regret bound with explicit and implicit cost estimations when $\beta = 1$. For the simulation, base learning rate parameters are set as in \cite[Theorems 3.1 and 3.3]{Bubeck2012} for Exp3 and Exp3P and in \cite[Theorems 1]{Neu2015} for Exp3IX and this work.

	Fig.~\ref{numerical2:regret1} shows that the proposed algorithm, a quantum bandit (QB), learns much faster and achieves better balancing of ExR/ExT searches without exploring the sub-optimal actions in an adversarial environment, compared to the counterparts. This is because QAA process associated with implicit exploration-style cost estimates allows to simultaneously amplify/attenuate the probabilities smoothly yielded from the learning scores, thus reducing the average regret by 50\% and 40\% from those of Exp3IX and QB with a sole ratio tune case ($\varrho > 1$, $\varsigma = 1$) requiring re-normalization \cite{dong2021}. Fig.~\ref{numerical2:regret2} demonstrates that the superior performance of the proposed algorithm is valid for different numbers of SPs $K$. A fine-grained implicit exploration approach could achieve higher and more robust performance, obtaining lower empirical mean and standard deviation of the regret than others.

	Fig.~\ref{numerical2:fig_ratio2} depicts the corresponding solution behaviors of $\varsigma$ and $\phi$ w.r.t $p_m$. i) The probability of {a dominant action increases} alongside the learning progress. A larger gap of probabilities between the dominant action and overall dominated actions, $p_m^t$ and $\frac{\sum_{k\in\mathcal{K}\backslash m}p_{k}^t}{K-1}$ guides us to set a lower $\phi$ (Prop.~\ref{prop_phi}). ii) As $K$ increases, the selected action $m$ with a given $p_m^t$ has higher dominance than the others,  $p_m^t\gg\frac{1-p_{m}^t}{K-1}$, and thus the chosen $\phi$ becomes lower, resulting in larger variability of $\phi$. iii) Meanwhile, the minimum limit of $\phi$ increases starting from $p_m^t$ equal to $\frac{1}{4}$ by Eq. (\ref{eq_range}) and the probability gap proportionally relative to the reduced range of $\phi$ yields the larger $\phi$. Choosing an appropriate value of $\phi \neq 0$ allows for simultaneously amplifying the amplitude of a dominant action while attenuating the ones of the others, thereby leading to better performance (Prop.~\ref{prop_regret}).
	
	The proposed algorithm has the potential for powerful computation in complex unknown environments, leveraging related quantum apparatuses. The quantum-inspired bandit algorithm is designed for quantum computers and motivated by quantum mechanics, but it is effective on traditional computers as well. This is due to two key aspects: (i) the collapse action selection strategy uses quantum measurement postulates to balance ExR-ExT trade-offs, without relying on empirical exploration parameter settings, and (ii) the probability magnitude updating strategy leverages quantum-mechanical phase control to simultaneously boost/suppress learning strategies based on the learning score, following the quantum superposition principle and without requiring additional normalization.

	\section{Conclusion}
	%In this work, we 
 This work proposed a quantum-inspired bandit learning algorithm to reduce the service cost under an adversarial environment. The proposed QAA approach allows for the new action update strategy and novel probabilistic action selection, provoked by the amplitude amplification and collapse postulate in quantum computation theory, respectively, together with a devised mapping between a quantum-mechanical phase in a quantum domain, and a distilled probability-magnitude in a value-based decision-making domain. This method effectively balances convergence speed and learning quality, outperforming traditional exploration approaches. Numerical results demonstrate its superiority over conventional methods.% in terms of regret performance.


\vspace{-.15cm}
	\begin{thebibliography}{100}		
		\bibitem{Mao2017}
		Y. Mao \emph{et~al.}, ``A survey on mobile edge computing: The communication perspective,'' \emph{IEEE Commun. Surveys Tuts.}, vol. 19, no. 4, pp. 2322-2358, Fourth quarter 2017.
		
%		\bibitem{Mouradian2018}
%		C. Mouradian \emph{et~al.}, ``A comprehensive survey on fog computing: State-of-the-art and research challenges,'' \emph{IEEE Commun. Surveys Tuts.}, vol. 20, no. 1, pp. 416-464, First quarter 2018.
		
		\bibitem{Cho2021}
		B. Cho \emph{et~al.}, ``Learning-based decentralized offloading decision making in an adversarial environment,'' \emph{IEEE Trans. Veh. Technol.}, vol. 70, no. 11, pp. 11308-11323, 2021.
		
		\bibitem{Kaelbling1993}	
		L. P. Kaelbling, ``Learning in Embedded Systems,'' \emph{MIT Press}, 1993.
		
		\bibitem{Auer2002}	
		P. Auer \emph{et~al.}, ``The nonstochastic multiarmed bandit problem,'' \emph{SIAM Journal on Computing}, vol. 32, no. 1, pp.48-77, 2002.
		 
		%
		\bibitem{cesa2017}	
		N. Cesa-Bianchi \emph{et~al.}, ``Boltzmann exploration done right.'' \emph{Adv. Neural Inf. Process. Syst.}, vol. 30, pp. 6284-6293, 2017.
		 
		\bibitem{Dunjko2018} 
		V. Dunjko \emph{et~al.}, ``Machine learning \& artificial intelligence in the quantum domain: a review of recent progress,'' \emph{Reports on Progress in Phys.}, no. 81, vol. 7, p.074001, 2018.
		
		
		\bibitem{Dong2008}
		D. Dong \emph{et~al.}, ``Quantum reinforcement learning,'' \emph{IEEE Trans. Syst. Man. Cybern. B}, vol. 38, no. 5, pp. 1207-1220, 2008.
		
		\bibitem{Grover} 
		L.~K.~Grover, ``Synthesis of quantum superpositions by quantum computation,'' \emph{Phys. Rev. Lett.}, vol. 85, pp.1334-1337, 2000. 
		
		\bibitem{Fakhari2013}
		P.~Fakhari \emph{et~al.}, ``Quantum inspired reinforcement learning in changing environment,'' \emph{New Math. and Natural Comput.}, vol. 9, no. 3, 2013.
		%%
		
		\bibitem{li2020}
		J. Li \emph{et~al.}, ``Quantum reinforcement learning during human decision-making,'' \emph{Nature Human Behaviour}, vol. 4, no. 3, pp. 294-307, 2020.
		
		\bibitem{dong2021}
		Y. Li \emph{et~al.}, ``Intelligent trajectory planning in UAV-mounted wireless networks: Quantum inspired reinforcement learning perspective,'' \emph{IEEE Wireless Commun. Lett.}, vol. 10, no. 9, pp.1994-1998, 2021.
		
		
			\bibitem{Casale2020}
		B. Casale \emph{et~al.}, ``Quantum bandits,'' \emph{Quantum Mach. Intell.}2, 2020.
		
		\bibitem{Wang2021} D. Wang \emph{et~al.}, ``Quantum exploration algorithms for multiarmed bandits'', \emph{Proc. the AAAI Conf. Artif. Intell.}, vol 35, no. 11, 2021.
		
		\bibitem{Wang2021_2} 
		D. Wang \emph{et~al.}, ``Quantum algorithms for reinforcement learning with a generative model'', \emph{Proc. Mach. Learn. Res.,} pp. 10916-10926, 2021.
		
		\bibitem{Rebentrost2021} 
		P. Rebentrost \emph{et~al.}, ``Quantum algorithms for hedging and the learning of ising models'', \emph{Phys. Rev. A}, vol. 103, p. 012418, 2021.
		
		
		
		\bibitem{Neu2015}
		G. Neu, ``Explore no more: Improved high-probability regret bounds for non-stochastic bandit,'' \emph{Adv Neural Inf Process Syst.}, vol. 28, 2015.
		
%		Neu, G., 2015. Explore no more: Improved high-probability regret bounds for non-stochastic bandits. Advances in Neural Information Processing Systems, 28.
		
%		\bibitem{cesa2006}
%		N. Cesa-Bianchi \emph{et~al.}, \emph{Prediction, learning, and games}, {Cambridge University Press}, 2006.
		
		\bibitem{Long2002}
		G. Long \emph{et.al}, ``Phase matching condition for quantum search with a generalized initial state,'' \emph{Phys. Lett. A}, vol. 294, pp.143-152, 2002.
		
		
		\bibitem{DHH} 
		C. Durr \emph{et.al}, ``A quantum algorithm for finding the minimum,'' arXiv preprint quant-ph/9607014, 1996.
		
		
		\bibitem{Bubeck2012}	
		S. Bubeck \emph{et~al.}, ``Regret analysis of stochastic and nonstochastic multi-armed bandit problems,'' \emph{Found. Trends in Mach. Learn.}, vol. 5, no. 1, pp. 1–122, 2012.
		
		
	 	\bibitem{Auer2002-2}
	 	P. Auer \emph{et~al.}, ``Finite-time analysis of the multiarmed bandit problem,'' \emph{Mach. Learn.}, vol. 47, no. 2, pp. 235-256, 2002.
		
	\end{thebibliography}
\end{document}